\documentstyle[prl,aps,multicol,epsf,rotate]{revtex}
\begin{document}

\draft
%\date{\today}
\title{Dramatic role of critical current anisotropy on flux avalanches in MgB$_2$ films}

\author{J. Albrecht, A. T. Matveev\cite{0}}
\address{Max-Planck-Institut f\"ur Metallforschung, Heisenbergstr. 3,
D-70569 Stuttgart, Germany}
\author{J. Strempfer, H--U. Habermeier}
\address{Max-Planck-Institut f\"ur Festk\"orperforschung, Heisenbergstr. 1,
D-70569 Stuttgart, Germany}
\author{D. V. Shantsev, Y. M. Galperin and T. H. Johansen}
\address{Department of Physics, University of Oslo, P. O. Box 1048,
Blindern, 0316 Oslo, Norway}
%\\ A. F. Ioffe Physico-Technical Institute, Polytekhnicheskaya
%26, St.Petersburg 194021, Russia}

\maketitle

%\pacs{74.70.Ad}%{Metals, alloys and binary compounds}
%\pacs{74.25.Qt}%{Vortex lattices, flux pinning, flux creep}
%\pacs{74.78.-w}%{Superconducting films and low-dimensional structures}

\begin{abstract}
Anisotropic penetration of magnetic flux in MgB$_2$ films grown on vicinal sapphire substrates is
investigated using magneto-optical imaging.
Regular penetration above 10~K proceeds more easily 
along the
substrate surface steps, anisotropy of the
critical current being 6\%.  
At lower temperatures the penetration occurs via abrupt dendritic avalanches that
preferentially propagate {\em perpendicular} to the surface steps. 
This inverse anisotropy in the penetration pattern becomes dramatic very close to 10~K 
where all flux avalanches propagate 
in the strongest-pinning direction. 
The observed behavior is fully explained using a thermomagnetic model
of the dendritic instability.
\end{abstract}

\begin{multicols}{2}
\narrowtext
\sloppy

Above the lower critical field $H_{c1}$ magnetic flux penetrates
type-II superconductors in the form of quantized vortices.
Usually, this penetration takes place  as a
regular flow leading to a smooth flux front\cite{Joos02}. However,
in some  superconducting films, e.g. of
MgB$_2$ a dramatic flux instability can occur 
resulting in an abrupt formation of magnetic
dendrites \cite{Joha02}. The flux
instability is believed to arise because the
motion of flux lines dissipates energy leading to local heating,
which results in a reduction of pinning and an enhancement of flux
line mobility. This constitutes a positive
feedback loop which acts as driving force for the
avalanches\cite{Mint81,Alts04}. Understanding and controlling
the avalanches is important not only from the fundamental point of
view, but also for applications since they represent huge
magnetic noise and  reduce the effective
critical current density.

Recently, numerous experimental investigations have
demonstrated that the dendritic instability is sensitive to
external conditions. The instability was found to disappear above
threshold values of temperature \cite{Joha02}, applied magnetic
field \cite{Rudn05}, and also for sufficiently small sample
dimensions\cite{Deni06}. Also covering MgB$_2$ films with a layer
of gold or aluminium was seen to suppress the dendrite formation
\cite{Bazi02} or change their propagation direction\cite{Albr05}.

To what extent the dendritic instability is sensitive to the
internal structure and flux pinning properties of the
superconducting film is an important and yet open question. A
recent thermomagnetic model predicts that pinning strength
represented by the critical current density $j_c$ controls the
onset of the dendritic instability \cite{Denisov06,Aranson05}. In
this paper we show that MgB$_2$ films with slightly anisotropic
$j_c$ can exhibit large anisotropy in the avalanche activity, in
full agreement with the model. The result gives direct evidence
that the microstructure of chemically homogeneous films grown on
vicinal substrates defines a direction for dendritic flux
propagation, which surprisingly turns out to be perpendicular to
the lowest pinning direction.

\begin{figure}[bbbb]
\begin{minipage}[b]{.24\textwidth}
\epsfxsize=1.\textwidth
\centerline{\epsffile{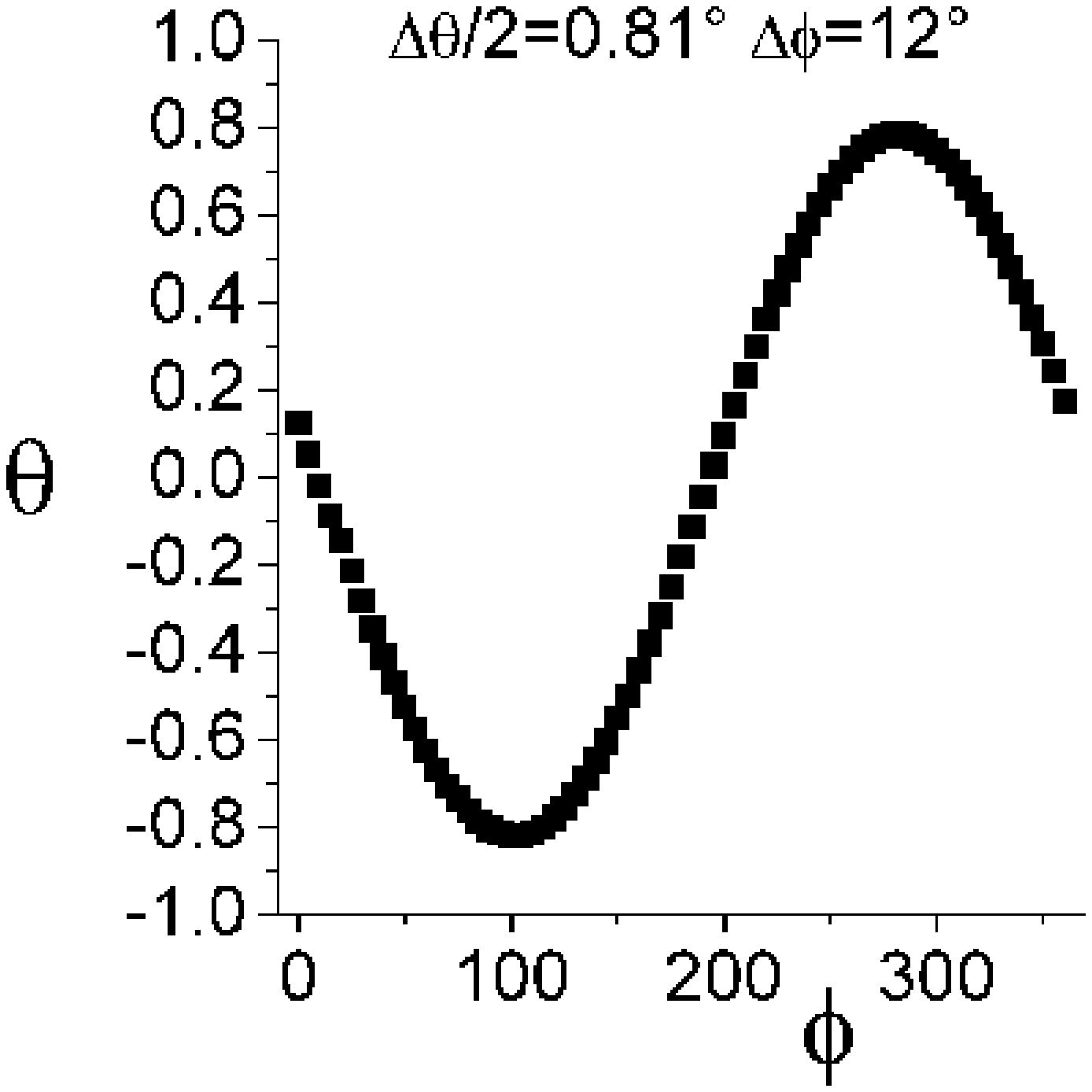}}
\end{minipage}
\hfill
\begin{minipage}[b]{.2\textwidth}
\epsfxsize=.9\textwidth \centerline{\epsffile{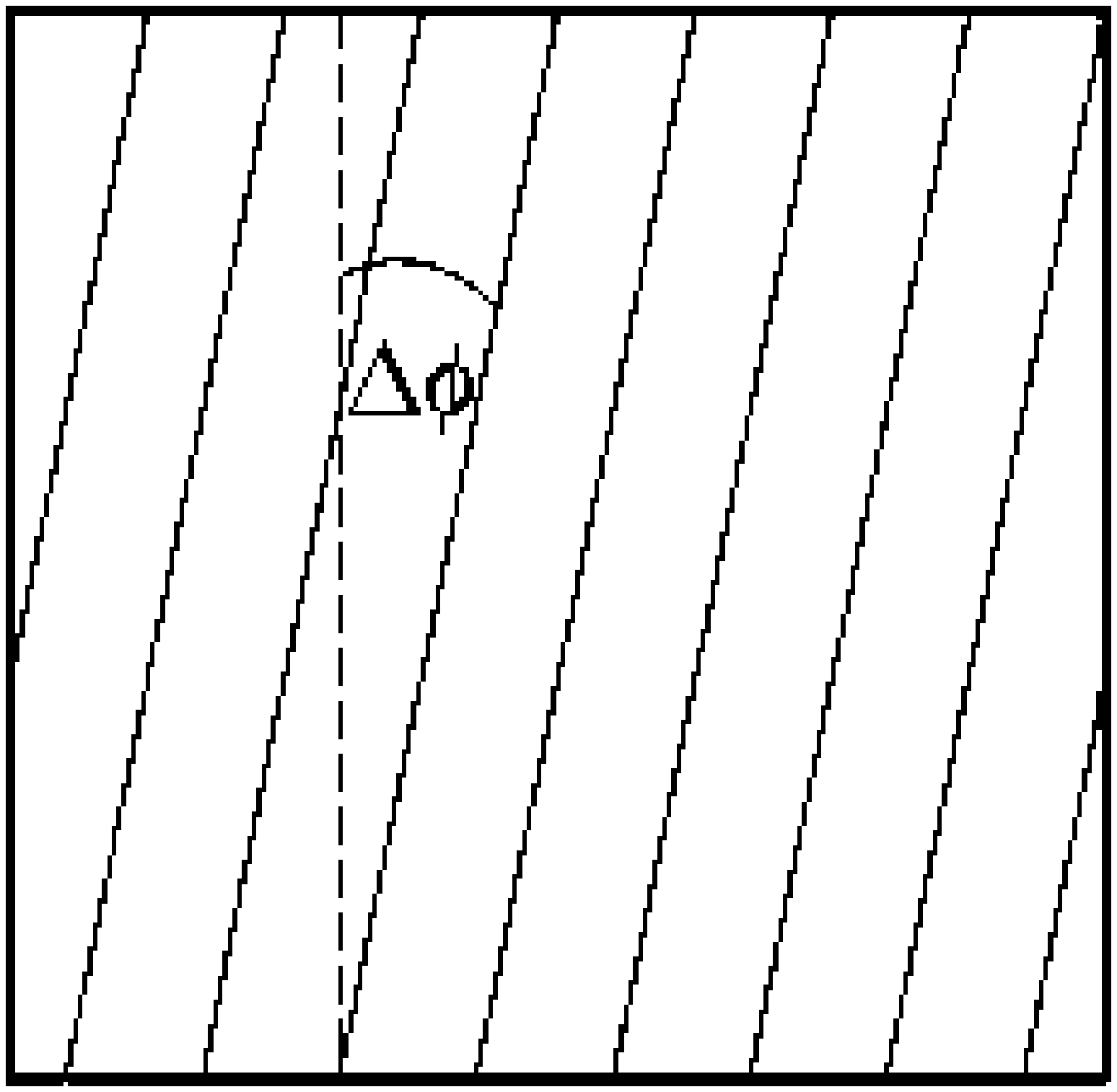}}
\epsfxsize=.9\textwidth \centerline{\epsffile{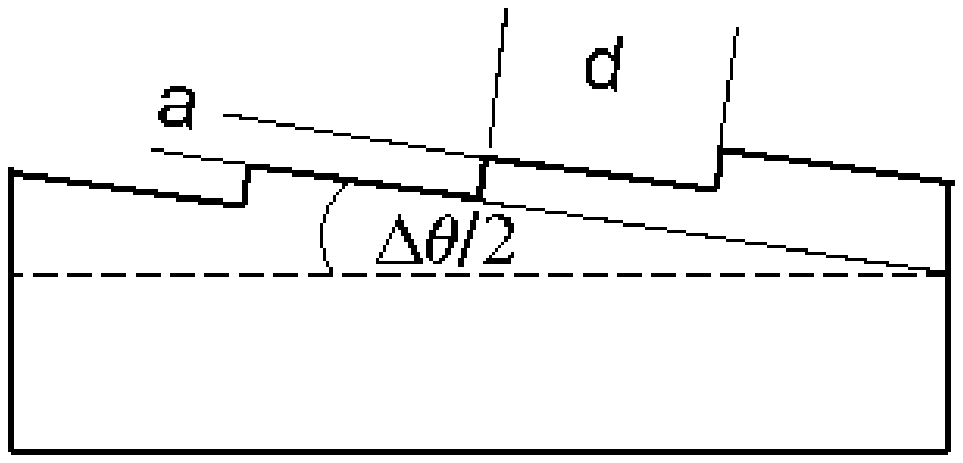}}
\end{minipage}\\

\caption{(Left) X-ray diffraction data for a sapphire substrate
slightly miscut with respect to the \mbox{[1~-1~0~2]} direction.
 (Right) Sketch of the square
substrate with surface steps defining the angles $\Delta \theta$
and $\Delta \phi$ which are determined from the plotted data.}
\label{xray}
\end{figure}

Substrate-induced modifications of 
the pinning properties 
%of thin films 
have successfully been
obtained previously in high-temperature superconducting films
grown on vicinal substrate surfaces\cite{Joos00,Djup05}, and
on artificially patterned substrates\cite{Albr04,Bruc05}. To
create anisotropic pinning in films of MgB$_2$ we used r-cut Al$_2$O$_3$
substrates slightly miscut relative to the \mbox{[1~-1~0~2]}
direction.
The substrate then
possesses a surface step structure   which can be
characterized by two angles $\Delta\theta$ and $\Delta\phi$,
according to the sketch in Fig.~\ref{xray}.
 Shown in the graph is the result of
x-ray diffraction measurements obtained from surface reflection of
a substrate mounted parallel to the $\phi$-axis of a 4-circle
diffractometer.   The deviation $\theta$ of the
direction of the \mbox{[1~-1~0~2]} reflection from the bisecting position
is plotted as function of the
 angle $\phi$. 
From the curve it follows that the substrate misorientation
angles are $\Delta\theta/2 = 0.81^\circ$ and $\Delta\phi =
12^\circ$. With a step height of one unit cell $a$ of Al$_2$O$_3$,
the step distance becomes $d = a/\tan\left(\Delta\theta/2\right) \approx$ 27~nm.

\begin{figure}[hhhht]
\epsfxsize=.3\textwidth \centerline{\epsffile{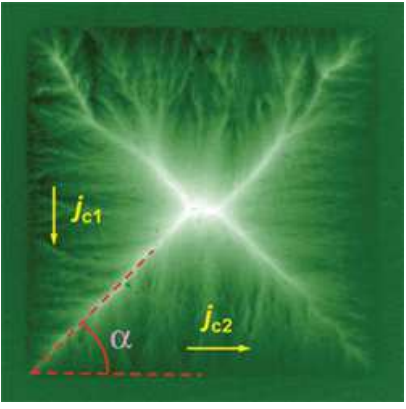}}
\vspace{4mm} \caption{Magneto-optical image of the remanent state
in an anisotropic $5\times 5$~mm$^2$ MgB$_2$ film at 12~K. The
angle $\alpha$ indicates the orientation of the discontinuity line
which allows determination of the critical current density ratio
$j_{c1}/j_{c2}$.} \label{rema12}
\end{figure}

The MgB$_2$ films were grown by sequential deposition of magnesium and boron layers
using a conventional electron beam evaporation and a subsequent
annealing process\cite{Shin01,Matv05}. The films selected for this
study had a lateral size of $5\times 5$~mm$^2$ and a thickness of
200~nm. The magnetic flux penetration into these 
films was investigated using the magneto-optical
(MO) Faraday effect allowing direct visualization of
the  flux  distribution\cite{Joos02}.
In our images the brightness represents the magnitude of the
flux density. 

To measure the anisotropy of the critical current density the
sample was first cooled to 12 K, and then exposed to a large
magnetic field which was subsequently removed. In this remanent
state, seen in Fig. \ref{rema12}, the flux distribution develops a
set of bright discontinuity lines (d-lines), where the current
sharply bends to adapt to the shape of the sample\cite{dlines}. In
an isotropic film the angle between the d-lines and the edge would
be $45^\circ$ and the d-lines would meet in the center of the
square. In our samples this is not the case, as seen directly by
the presence of a horizontal bright line segment in the central
part of the square. Evidently, the current flowing parallel to the
vertical edges gives stronger shielding than the current flowing
horizontally. Assuming Bean-model behavior, the two critical
current densities $j_{c1}$ and $j_{c2}$ are related to the d-line
angle $\alpha$, see the figure, by $j_{c1}/j_{c2} = \tan \alpha$.
Measurement of the angle gives $\tan \alpha = 1.06$, i.e., a 6\%
anisotropy. Comparing with the direction of the steps on the
substrate surface, see Fig. 1, the larger value of $j_{c1}$
represents a pinning force enhancement for flux moving
perpendicular to the surface steps.

\begin{figure}[hhht]
\hfill
\begin{minipage}[b]{.3\textwidth}
\epsfxsize=1.\textwidth \centerline{\epsffile{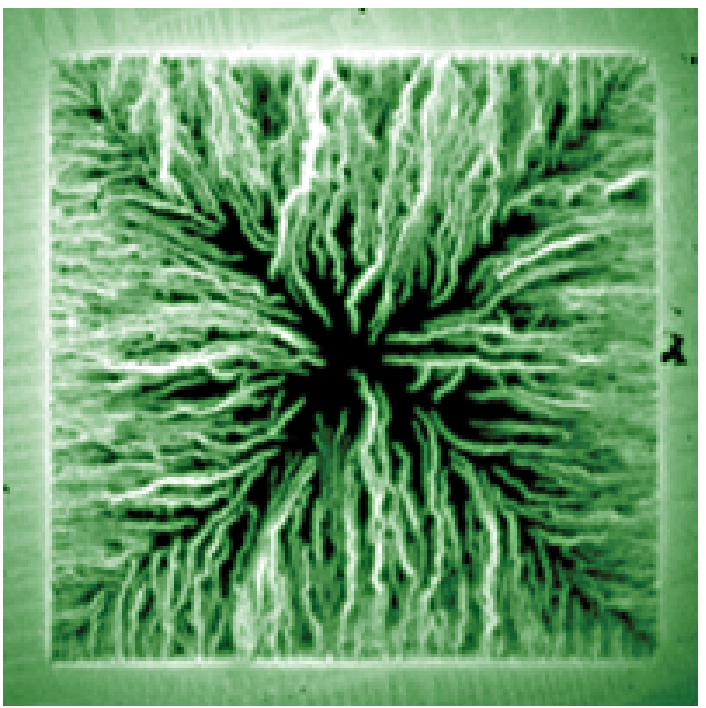}}
\epsfxsize=1.\textwidth \centerline{\epsffile{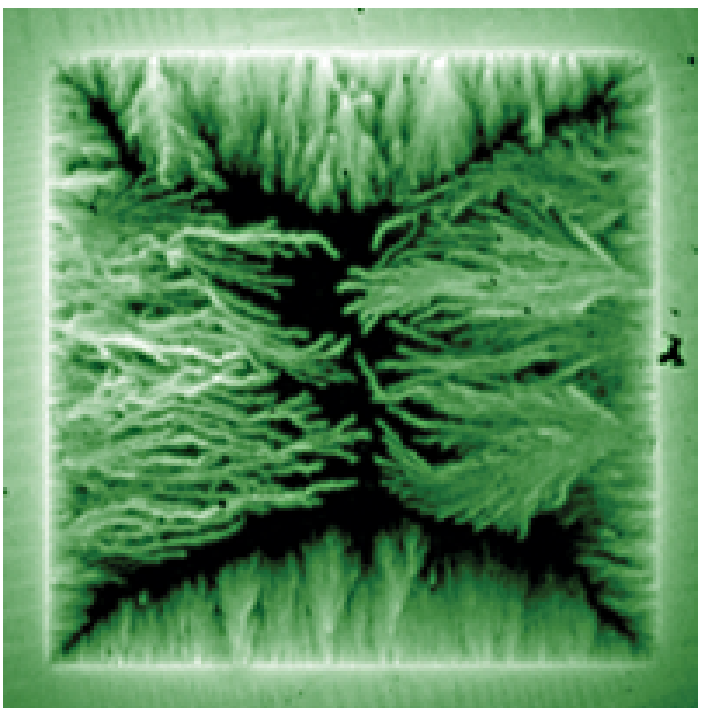}}
\end{minipage}
\begin{minipage}[b]{.1\textwidth}
~$T=8$ K \\[46mm]

~$ T=10$ K\\
\end{minipage}\\
\caption{ Flux penetration into the same MgB$_2$ film as in
Fig.~\ref{rema12}, but at lower temperatures and an applied field
of 16~mT. At 8 K the flux pattern is dendritic and isotropic, while
at 10~K dendritic avalanches are formed only along one direction.}
\label{aniso}
\end{figure}

Shown in Fig.~\ref{aniso} is the flux distribution after
applying a 8 mT magnetic field to the MgB$_2$ film initially
zero-field-cooled to the slightly lower temperatures. Here one
should expect the flux to penetrate in the form of dendritic
avalanches\cite{Joha02}, which indeed is what we observe. At 8 K
the avalanches are seen to enter the film evenly from all the
edges. At
10 K, however, the penetration pattern suddenly becomes strongly
anisotropic; flux dendrites develop from the vertical edges only,
whereas from the top and bottom edges the penetration is smooth.
In other words, the flux dendrites propagate predominantly in the
direction perpendicular to the substrate surface steps, i.e., the
direction of the stronger flux pinning. It is indeed highly
counterintuitive that the flux prefers to move perpendicular
rather than parallel to the linear defects introduced in the
sample.

\begin{figure}%[hhhhhh]
\hfill
\begin{minipage}[b]{.3\textwidth}
\epsfxsize=1.\textwidth \centerline{\epsffile{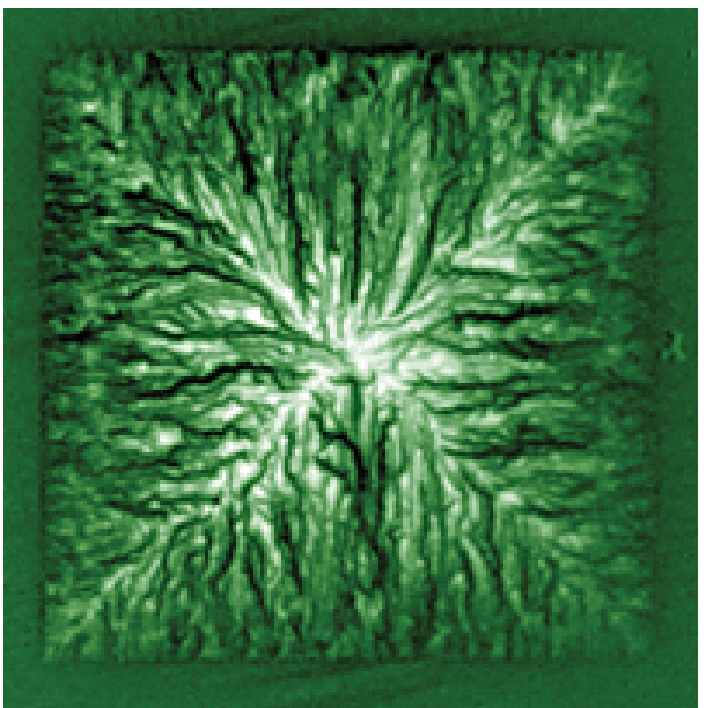}}
\epsfxsize=1.\textwidth \centerline{\epsffile{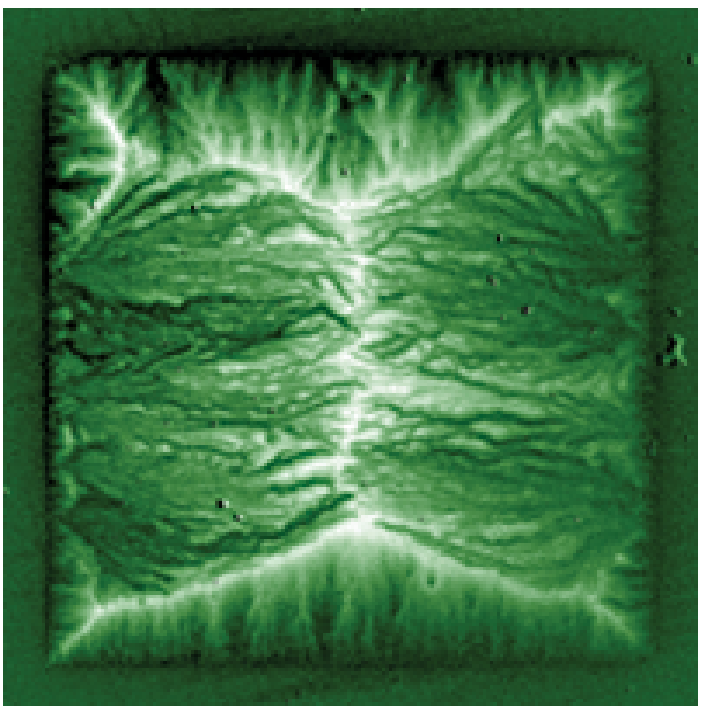}}
\end{minipage}
\begin{minipage}[b]{.1\textwidth}
~$T=8$ K \\[46mm]

~$ T=10$ K\\
\end{minipage}\\

\vspace{4mm} \caption{Remanent states of the anisotropic MgB$_2$
film. The flux pattern is isotropic at 8~K, but strongly
anisotropic at 10~K, similar to the virgin penetration behavior in
Fig.~\ref{aniso}.} \label{rema}
\end{figure}

A quite similar behavior was observed for the penetration
behavior during field descent, see Fig. \ref{rema}. The images
show the remanent state after applying a maximum field of
$\mu_0H_{}= 0.2$~T. Again at 8 K, the penetration is isotropic
only now the dendrites appear dark since they consist of antiflux
surrounded by annihilation zones. At 10 K the dendrites propagate
also now along the horizontal direction dictated by the
microstructural anisotropy.

%\end{widetext}

All these experiments show the remarkable fact that an anisotropy
in $j_c$ as small as 6\% can lead to a dramatic anisotropy in the
avalanche activity near 10 K, and only very near this temperature.
This behavior can be understood using results of a recent model of
the dendritic instability developed in
Refs.~\onlinecite{Denisov06,Aranson05}. The model is based on
stability analysis of the thermal diffusion and Maxwell equations
in a long and thin superconducting strip thermally coupled to a
substrate. Initially, the strip is placed in an increasing
perpendicular magnetic field, and a Bean-like critical state is
formed in the flux penetrated region. This state can become
unstable with respect to perturbations in the magnetic field and
temperature, and under some conditions a fastest growing
perturbation has a non-zero wave vector along the film edge. This
means that an instability will develop in the form of narrow
fingers perpendicular to the edge -- a scenario closely resembling
the observed dendritic flux behavior. The threshold flux
penetration depth, $\ell^*$, when the superconducting strip first
becomes unstable, is given by Eq.~(25) of Ref. \cite{Denisov06},
which can be written as

\begin{equation}
\ell^* =\frac{\pi}{2} \sqrt{\frac{\kappa T^*}{j_c E}}
\left( 1 - \sqrt{\frac{2h_0 T^*}{nd j_c E}}\right)^{-1}\, .
\label{kxy}
\end{equation}
Here, $d$ is the film thickness, $T^*\equiv -(\partial \ln
j_c/\partial T)^{-1}$, $\kappa$~is the thermal conductivity, and
$h_0$ is the coefficient of heat transfer from the superconducting
film to the substrate.
%, i. e., the boundary condition at the interface
%is $\kappa \nabla T = h_0 (T-T_0)$, where $T_0$ is the substrate
%temperature.
The parameter $n$ characterizes the nonlinearity of the
current-voltage curve of the superconductor, $n=\partial \ln
E/\partial \ln j \gg 1$.

The threshold field, $H_{th}$, when the first dendrite invades
the film is obtained by combining (\ref{kxy}) with the Bean model expression,
\begin{equation}
%\frac{\ell}{w}= 1- \frac{1}{\cosh \left({\pi H}/{ j_c d}\right)}
H =  \frac{j_c d}{\pi} \;  {\rm arccosh} \left(\frac{w}{w-\ell}
\right) ,
\label{lH}
\end{equation}
which connects the the applied field with the flux penetration
depth $\ell$ for a long thin strip of width $2w$~\cite{BrIn}.

\begin{figure}%[hhht]
\epsfxsize=.5 \textwidth \centerline{\epsffile{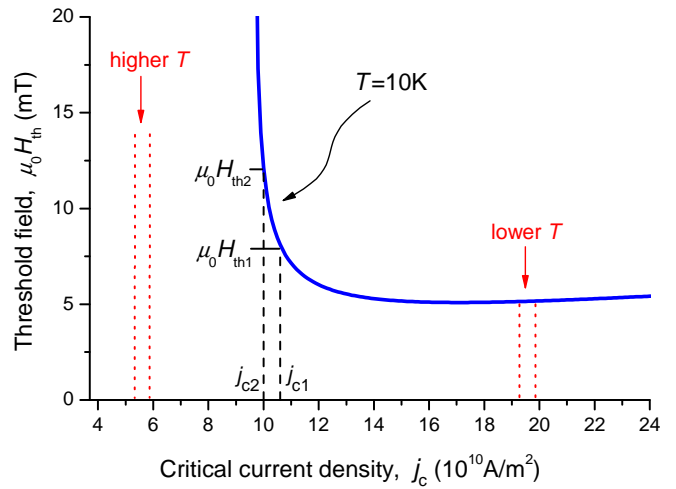}}
%\centerline{\includegraphics[width=8cm]{jc.eps}}
\caption{Threshold field for the dendritic instability as a function of the
critical current density, calculated using (\ref{kxy}) and (\ref{lH}). Two
dashed lines show the values of $j_c$ in an MgB$_2$ film at 10~K for the
directions along and across the substrate steps. Large difference in the
corresponding $H_{th}$'s  implies a strong anisotropy of the dendritic
avalanches specifically at $T=10$~K.}
\label{Jc}
\end{figure}

Fig.~\ref{Jc} shows the dependence of $H_{th}$ on $j_c$ calculated
using Eqs. (\ref{kxy}) and (\ref{lH}) assuming $\kappa
T^*/E=140$~A and $h_0 T^*/nE=9230$~A/m (which can mean e.g.
$T^*=10$~K, $E=10$~mV/m, $\kappa=0.14$~W/Km, $n=30$, and
$h_0=280$~W/Km$^2$). The graph shows that for high critical
currents the threshold field is weakly dependent on $j_c$.
However, when $j_c$ becomes smaller, the expression in the
brackets of Eq. (\ref{kxy}) approaches zero, and $H_{th}$
diverges. When $j_c$ drops below a certain value, the dendritic
instability is absent no matter how large field is applied.

The critical current density at 10~K can be estimated using Eq.
(\ref{lH}) from the flux penetration depth just before the
avalanche behavior sets in.
We then find
$j_{c2} =1\times 10^{11}$~A/m$^2$, and $j_{c1}$ larger by 6\%.
These critical current values found from MO images agree with
results obtained by SQUID measurements \cite{Matv05}. The two
values for $j_c$ are indicated by two dashed lines in
Fig.~\ref{Jc}. Despite the small difference between $j_{c1}$ and
$j_{c2}$, the corresponding threshold fields $H_{th1}$ and
$H_{th2}$ differ significantly. When the increasing applied field
reaches the lowest of the two threshold fields, $H_{th1}$,  the
dendrites should appear from the sides where $j_c$ is the highest.
This is exactly what one can see from Fig.~\ref{aniso} (bottom).
Interestingly, this penetration pattern dominated by dendrites
shows an inverted anisotropy compared to the critical state
pattern in Fig.~\ref{rema12}.

As the applied field is further increased and reaches $H_{th2}$,
one could expect that dendrites appear also from the top and
bottom edges. However, this does not happen experimentally, as can
be seen from the 10 K images. The reason probably is the dramatic
disturbance in the current flow created by the dendritic
structures formed at the smaller fields. They fill almost the
whole interior of the film, thus lowering the edge fields
substantially, which prevents invasion of new dendrites.

At a higher temperature, the critical current density becomes
smaller, i.e., we ``move'' to the left along the $x$ axis in
Fig. \ref{Jc}, and enter the stable region. This is again in full
agreement with experiment, where we do not observe dendritic
avalanches above 10~K. Conversely, by lowering the temperature we
move to the right, where $H_{th}$ is almost independent of $j_c$.
Hence, the dendritic avalanches should penetrate evenly from all
four sides of the square film. This is exactly what we find at
8~K, see Fig.~\ref{rema}~(top), and at lower $T$.\cite{footnote}

Note, that the $H_{th}(j_c)$ curve is slowly rising at large
$j_c$. It means that at very low $T$ one should expect the
opposite kind of anisotropy in the dendritic growth. Namely, the
dendrites should first appear from the sides where the regular
flux penetration is {\em deeper}. However, in order to observe
this effect one  needs a sample where the anisotropy in $j_c$ is
much stronger than the 6\% in the present case.

The analysis above shows that the dendritic penetration has a
pronounced anisotropy only in the narrow interval of temperatures
corresponding to the steep region at the $H_{th}(j_c)$ curve. For
other superconducting samples this region will not necessarily be
located at 10~K, but it will always be near the threshold
temperature above which the instability disappears.

In conclusion, we have investigated the magnetic flux penetration
into MgB$_2$ films grown on vicinal sapphire substrates. The
resulting microstructure giving a slight anisotropy in $j_c$ leads
to a large anisotropy in the flux patterns, but only in a narrow
temperature interval near 10~K. Here, the dendritic avalanches
propagate only in the direction perpendicular to the substrate
surface steps, i.e., the flux moves along the direction of {\em
largest} pinning. The surprising behavior is fully explained by a
thermo-magnetic model for the dendritic instability.

%Our results provide a clear proof that the formation of dendritic
%avalanches is sensitive to local defect structures and, in
%particular, increasing $j_c$ makes films more unstable.

%{\references

\end{multicols}

\begin{thebibliography}{99}
\bibitem{footnote}  In principle, one should also take into account
that $\kappa$ and $h_0$ are expected to increase with temperature.
It can be easily verified that for $T>10$~K this shifts the
$H_{th}(j_c)$ curve to the right, while for $T<10$~K -- to the
left. This produces the same relative shifts as the $T$-dependence
of $j_c$, and hence
 does not change the conclusions of the previous paragraph.
\end{thebibliography}
\end{document}